\documentstyle[12pt]{article}

\let\a=\alpha    \let\b=\beta         \let\d=\delta
     
            \let\q=\theta       
 \let\k=\kappa
\let\l=\lambda  \let\m=\mu                  \let\x=\xi
 \let\p=\pi
          \let\t=\tau            
   \let\f=\phi
\let\c=\chi                      
\let\w=\omega

\let\X=\Xi                        
 
\let\Y=\Psi                   \let\L=\Lambda

      \let\pa=\partial      \let\bm=\bibitem
\newcommand{\be}{\begin{equation}}
\newcommand{\ee}[1]{\label{#1}\end{equation}}
\newcommand{\bea}{\begin{eqnarray}}
\newcommand{\eea}{\end{eqnarray}}
\newcommand{\ra}{\rightarrow}

\newcommand{\dV}[1]{\frac{\d}{\d V(#1)}}
\newcommand{\dY}[1]{\frac{\d}{\d \Y (#1)}}
\newcommand{\sect}[1]{\setcounter{equation}{0} \section{#1}}

\newcommand{\refer}[1]{(\ref{#1})}
\newcommand{\Blangle}{\Bigl \langle}
\newcommand{\Brangle}{\Bigr \rangle}
\newcommand{\cint}[2]{\oint_{#2} \, \frac{d #1}{2\p i}}
\newcommand{\XL}{( \X_1 - \L_1 )}
\newcommand{\Yop}{\Bigl (\, \widehat{\Y} \, -\, v_0(p)\, \Bigr )}
\newcommand{\Vop}{\Bigl (\, \widehat{V}^\prime \, -\, u_0(p)\, \Bigr )}

\newcommand{\Fbg}{F^{\mbox{\scriptsize bos}}_g}
\newcommand{\Ffg}{F^{\mbox{\scriptsize ferm}}_g}
\newcommand{\dol}[1]{\mbox{$#1$}}

\begin{document}
\thispagestyle{empty}
\renewcommand{\thefootnote}{\fnsymbol{footnote}}
{\hbox to\hsize{
\vbox{\noindent ITP--UH--14--95 \hfill April 1995 \\
hep-th/9504089}}}
\noindent
\vskip2cm
\begin{center}

{\Large\bf The Supereigenvalue Model in the} \vglue .2cm
{\Large\bf  Double--Scaling Limit}
\vglue2cm

{\sc Jan C.\ Plef\/ka }\footnote[2]{Supported by
the `Studienstiftung des Deutschen Volkes'}
\vglue1cm
{\it Institut f\"{u}r Theoretische Physik, Universit\"{a}t Hannover}\\
{\it Appelstra\ss{}e 2, 30167 Hannover, Germany}\\
{\footnotesize plefka@itp.uni-hannover.de}

\vglue2cm
{ABSTRACT}
\end{center}
\noindent
The double--scaling limit of the supereigenvalue model is performed in the
moment
description. This description proves extremely useful for the identification
of the multi-critical points in the space of bosonic and fermionic
coupling constants. An iterative procedure for the calculation of higher--genus
contributions to the free energy and to the multi--loop correlators
in the double--scaling limit is developed.
We present the general structure of these quantities
at genus $g$  and give explicit results up to and including genus two.
\vfill
\renewcommand{\thefootnote}{\arabic{footnote}}
\setcounter{footnote}{0}
\setcounter{page}{0}
\pagebreak

\sect{Introduction}

To date the most promising {\it discrete} approach to 2D supergravity coupled
to
minimal superconformal models is supplied by the supereigenvalue model
proposed by Alvarez--Gaum\'e {\it et  al.\ }\cite{Alv}. It is formulated  in
terms
of a collection of $N$ Grassmann even and odd variables, the
``supereigenvalues'',
as well as a set of even and odd coupling constants. For a detailed review
see ref.\ \cite{Book}.
There still is no progress towards an understanding of the model on the basis
of a generalized matrix model, which might provide us with
 a geometrical picture in the form of discretized super--Riemann surfaces.
\footnote{The only attempt in this direction is ref. \cite{Tak}}
Nevertheless many of the well
known features of the hermitian matrix model, such as the genus expansion,
the Virasoro constraints, the loop equations, the moment description and the
loop insertion operators \cite{Amb}, find their supersymmetric counterparts in
the supereigenvalue model. From this point of view the supereigenvalue model
appears as the natural supersymmetric generalization of the hermitian
one--matrix model.

The supereigenvalue model is solvable nonperturbatively in coupling constants
but perturbatively in its genus expansion. The solution is based on a set of
superloop equations obeyed by the superloop correlators. Away from the double
scaling limit these equations were first solved for general potentials in the
planar limit in ref.\ \cite{Alv2}. In the moment description
the computation of higher genus contributions could be automatized
by an iterative procedure yielding results for general potentials and
in principle arbitrary genus \cite{Ple}, representing a generalization of this
very effective
method for the hermitian matrix model \cite{Amb}. Explicit results were given
for genus one. An alternative approach was pursued by the authors of refs.\
\cite{Bec,McA}
who managed to directly integrate out the Grassmann--odd variables on the level
of the partition function, uncovering the maximally quadratic dependence of
the free energy on the fermionic coupling constants. The supereigenvalue
model also displays a connection to supersymmetric integrable models
\cite{FigMcA2}.

In order to make contact with continuum physics the supereigenvalue
model has to be studied in its double--scaling limit. This was done for
symmetric
bosonic potentials in ref.\ \cite{Alv}in the planar limit, the case
of general potentials was
solved in ref.\ \cite{Alv2} for genus zero, one and partially for genus two. By
making use of these results the authors of ref.\ \cite{Zad} developed
a precise dictionary between continuum \dol{N=1} super--Liouville amplitudes
and supereigenvalue correlators.

In this paper the methods of the iterative solution of the supereigenvalue
model
\cite{Ple} are applied to the double--scaling limit. The moment
description turns out to be extremely useful for the determination of the
critical
points in the space of coupling constants. Just as in the hermitian matrix
model
Kazakov multicritical points \cite{Kaz} appear, related to extra zeros of the
eigenvalue densities accumulating at one endpoint of the support. We identify
the scaling properties of the moments and basis functions, confirming the
results of refs.\ \cite{Alv,Alv2}.
The iterative procedure may be optimized to only
produce terms relevant in the double--scaling limit. With these methods at hand
we are able to state the general structure of the double scaled superloop
correlators and the free energy. The iteration is perfectly suited for
implementation on a computer algebra system. Explicit results are stated up to
and including genus two.

The paper is organized as follows: In section 2 we briefly review the
supereigenvalue model, the superloop equations and their iterative solution.
In section 3 the double--scaling limit is performed. The multi--critical
points in the moment description are identified and the scaling behaviour of
moments and basis functions is obtained. We proceed to develop the iterative
process in the scaling limit, state the general structure of the superloop
correlators and the free energy and present our explicit results. Section 4
finally contains the conclusions and a short discussion of future perspectives.

\sect{Iterative Solution of the Supereigenvalue Model}

In the following we give a brief account of the iterative procedure to
solve the supereigenvalue model genus by genus for general potentials.
A more elaborate description may be found in ref.\ \cite{Ple}. The aim of this
paper
is to apply these methods to the double--scaling limit.

\subsection{Superloop Insertion Operators}

The supereigenvalue model \cite{Alv} is defined by the partition
function

\be
{\cal Z}= e^{N^2 \, F} =
\int (\prod_{i=1}^{N} d\l_i \, d\q_i ) \, \prod_{i<j}\, (\l_i -
\l_j - \q_i\q_j )
\,\exp \Bigl ( - N
\sum_{i=1}^N \, [ V(\l_i) - \q_i \Y (\l_i)] \, \Bigr ),
\ee{model}
where the integration is over a set of $N$ bosonic and fermionic variables,
denoted by $\l_i$ and $\q_i$ respectively. $N$ is even. Moreover the Grassmann
even and odd potentials take the general form

\be
V(\l_i)= \sum_{k=0}^\infty g_k {\l_i}^k   \quad \mbox{and} \quad \,\,\,
\Y(\l_i) = \sum_{k=0}^\infty \x_{k+1/2}  {\l_i}^k,
\ee{}
the $g_k$ and and $\x_{k+1/2}$ being Grassmann even and odd coupling
constants, respectively. The basic observables of this model are the
connected $(n|m)$--superloop correlators

\bea
\lefteqn{W(p_1,\ldots ,p_n \mid q_1,\ldots ,q_m)=}\label{Wnm}\\ & &
N^{n+m-2} \,
\Blangle \,
\sum_{i_1} \frac{\q_{i_1}}{p_1-\l_{i_1}} \, \ldots \, \sum_{i_n}
\frac{\q_{i_n}}{p_n-\l_{i_n}} \, \sum_{j_1}\frac{1}{q_1-\l_{j_1}} \,
\ldots\, \sum_{j_m} \frac{1}{q_m-\l_{j_m}} \,\Brangle _{
\mbox{\scriptsize conn}} ,
\nonumber\eea
where the expectation value is defined in the conventional way and where
``conn'' refers to the connected part. These superloop correlators act as
generating functionals for higher--point correlators like
\mbox{$\langle \sum_i \q_i {\l_i}^k \sum_j {\l_j}^l \rangle$} upon expansion
in  $p_i$ and $q_i$. The connected $(n|m)$--superloop correlators are
related to the free energy by the application of the superloop insertion
operators $\d /\d V(p)$ and $\d/\d\Y (p)$:

\be
W(p_1,\ldots ,p_n \mid q_1,\ldots ,q_m)=
\dY{p_1} \,\ldots\,
\dY{p_n}\, \dV{q_1}\,\ldots\, \dV{q_m}
\,\, F,
\ee{Wnm2}
where

\be
\dV{p}= -\sum_{k=0}^\infty\frac{1}{p^{k+1}} \frac{\pa}{\pa g_k}
 \quad\mbox{and}\quad
\dY{p}= -\sum_{k=0}^\infty\frac{1}{p^{k+1}} \frac{\pa}{\pa \x_{k+1/2}}.
\ee{superloopinsertionops}

Hence from the one--superloop correlators $W(\, |p)$ and $W(p|\, )$ (or the
free energy $F$) all the multi--superloop correlators can be obtained by
application of the superloop insertion operators.

With the normalizations chosen above, the genus expansion of the correlators
reads

\be
W(p_1,\ldots ,p_n \mid q_1,\ldots ,q_m)= \sum_{g=0}^\infty \, \frac{1}{N^{2g}}
\, W_g (p_1,\ldots ,p_n \mid q_1,\ldots ,q_m).
\ee{genusexpWnm}
Similarly we have

\be
F= \sum_{g=0}^\infty \, \frac{1}{N^{2g}} \, F_g
\ee{genusexpF}
for the free energy.

\subsection{Superloop Equations}

The genus $g$ contribution to the one--superloop correlators $W(\, |p)$ and
$W(p|\, )$ may be found by solving the superloop equations of the model by
iteration. These two equations express the invariance of the partition
function \refer{model} under
field redefinitions as shown in ref.\ \cite{Ple}.

The Grassmann--odd superloop equation takes the form

\be
\cint{\w}{C} \, \frac{V^\prime (\w)}{p-\w}\, W (\w \mid\, ) \, + \,
\cint{\w}{C} \, \frac{\Y (\w)}{p-\w} \, W (\, \mid \w) =
W(p \mid \, ) \, W(\,\mid p) \, + \, \frac{1}{N^2} \, W(p \mid p)
\ee{superloop1}
and its counterpart, the Grassmann--even superloop equation, reads

$$
\cint{\w}{C} \, \frac{ V^\prime (\w )}{p - \w} W(\, \mid \w ) \, + \,
\cint{\w}{C} \,\frac{\Y ^\prime (\w )}{p - \w} \,
W(\w \mid \, ) \, -\, \frac{1}{2} \, \frac{d}{dp}\, \cint{\w}{C} \,
\frac{\Y (\w )}{p-\w} \, W( \w \mid )\, = \phantom{W(p \mid \, )\,}$$
 \be
 \frac{1}{2} \, \Bigl [ \, W(\, \mid p)^2 \, - \, W(p \mid \, )\, W^\prime (p
 \mid \, ) \, + \, \frac{1}{N^2}\, \Bigl ( \, W(\, \mid p,p) \, - \,
\frac{d}{dq}\,
 W( p,q \mid \, ) \,  \Bigr | _{p=q} \, \Bigr ) \Bigr ] .
\ee{superloop2}
In the derivation we have assumed a one--cut structure of the loop correlators,
i.e.\ in the limit $N\ra \infty$ we assume that the eigenvalues are contained
in a finite interval \mbox{$[x,y]$}. $C$ is a curve around the cut.
 Moreover eqs.\ \refer{superloop1} and \refer{superloop2} encode the
super--Virasoro constraints $G_{k+1/2}\,
{\cal Z}=0$ and $L_k\, {\cal Z} = 0$ for $k \geq -1$, which the model obeys by
construction \cite{Alv}.
As demonstrated in ref.\ \cite{Ple} these equations are accessible to an
iterative
solution in genus $g$. The strategy consist in employing the observation of
refs.\
\cite{Bec,McA} that the free energy $F$ depends at most quadratically on
fermionic coupling constants. By eq.\ \refer{Wnm2} this allows us to write

\be
W(p \mid \, ) = v(p) \quad\mbox{and}\quad
W(\, \mid p)  = u(p) \, + \, \widehat{u}(p).
\ee{21314}
Here $v(p)$ is of order one, $u(p)$ of order zero and $\widehat{u} (p)$ of
order
two in the fermionic coupling constants $\x_{k+1/2}$. Plugging these relations
and the genus expansion \refer{genusexpWnm} into the superloop equations
\refer{superloop1} and \refer{superloop2} yields a set of four equations at
each
genus sorted by their order in fermionic coupling constants for the quantities
$v_g(p)$, $u_g(p)$ and $\widehat{u}_g(p)$.

The equation of order 0 in fermionic couplings at genus \mbox{$g\geq 1$} reads

\bea
\Vop \, u_g(p) &=&
\frac{1}{2}\, \sum_{g^\prime=1}^{g-1}\, u_{g^\prime}(p)\, u_{g-g^\prime}(p)
\, +\, \frac{1}{2}\, \dV{p}\, u_{g-1}(p)\, \nonumber \\ & &
-\, \frac{1}{2}\, \frac{d}{dq}\,\dY{p}\,
v_{g-1}(q)\Bigr | _{p=q},
\label{order0genusg}
\eea
where we have introduced the linear operator $\widehat{V}^\prime$ defined by

\be
\widehat{V}^\prime \, f(p) = \cint{\w}{C}\, \frac{V^\prime (\w )}{p-\w}\,
f(\w ).
\ee{Vprimedef}
The equation of order 1 at $g\geq 1$ takes the form

\bea
\Vop \, v_g(p)  &=& -
\Yop\, u_g(p)\, +\,  \sum_{g^\prime=1}^{g-1}\, v_{g^\prime}(p)
\, u_{g-g^\prime}(p) \, \nonumber \\ & &
+\, \dV{p}\, v_{g-1}(p),
\label{order1genusg}
\eea
analogously the operator $\widehat{\Y}$ is defined by

\be
\widehat{\Y}\, f(p) = \cint{\w}{C}\, \frac{\Y (\w )}{p-\w}\,
f(\w ).
\ee{opVprime}
In fact for $g\geq 1$ there is no need to solve the equations of order 2 and
3, as the knowledge of the quantities $u_g(p)$ and $v_g(p)$ (up to a zero
mode) suffices to compute the free energy $F_g$ and the remaining
$\widehat{u}_g$ \cite{Ple}. It is the remarkable structure of eqs.\
\refer{order0genusg} and \refer{order1genusg} which allows us to develop
an iterative procedure to solve for $u_g$ and $v_g$. At genus $g$ the right
hand side of \refer{order0genusg} only contains contributions from lower
genera, knowing these the inversion of the operator \dol{(\widehat{V}^\prime
-u_0(p))} then yields $u_g(p)$. Similarly
the right hand side of eq. \refer{order1genusg} contains $u_g$
and  terms of lower genera, allowing the computation of $v_g(p)$ (up to the
zero
mode of the operator \dol{(\widehat{V}^\prime
-u_0(p))} ). Hence the seed of the solution is $u_0$ and $v_0$.
Let us now turn to an important
tool of the solution, the change of variables from coupling constants to
moments and the introduction of basis functions.

\subsection{Moments and Basis Functions}

The planar as well as the higher genera contributions to the loop correlators
and the free energy are most easily expressed by introducing instead of the
couplings \mbox{$\{ g_k, \x_{k+1/2}\}$} a set of Grassmann even and odd
moments \mbox{$\{ M_k,J_k,\X_k,\L_k\}$}.
We replace the bosonic couplings $g_k$ by the moments \cite{Amb}

\bea
M_k &=& \cint{\w}{C}\, \frac{V^\prime (\w )}{ (\w - x)^k} \,
\frac{1}{[\, (\w -x)\, (\w-y)\, ]^{1/2}}
, \quad k\geq 1 \\ &&\nonumber\\
J_k &=& \cint{\w}{C}\, \frac{V^\prime (\w )}{ (\w - y)^k} \,
\frac{1}{[\, (\w -x)\, (\w-y)\, ]^{1/2}}
, \quad k\geq 1 ,
\eea
and the fermionic couplings $\x_{k+1/2}$  by

\bea
\X_k &=& \cint{\w}{C}
\, \frac{\Y (\w )}{(\w -x)^k}\, [\,(\w -x)(\w -y)\, ]^{1/2}
,\quad k\geq 1 \\ &&\nonumber\\
\L_k &=& \cint{\w}{C}\,
\frac{\Y (\w )}{(\w -y)^k}\, [\,(\w -x)(\w -y)\, ]^{1/2}
,\quad k\geq 1 .
\eea

One advantage of
this change of variables is that the contributions at genus \mbox{$g\geq 1$}
 to the loop
correlators and the free energy depend only on a finite number of moments,
e.g.\ $F_g$ depends at most on \mbox{$2 \times 3g$} bosonic and
\mbox{$2\times (3g+1)$} fermionic moments \cite{Ple}.
As opposed to this $F_g$ is a
function of the entire set of coupling constants \mbox{$\{g_k,\x_{k+1/2}\}$}.
Furthermore the description in terms of moments will prove extremely useful
in the double--scaling limit.

For the development of the iterative procedure we further introduce
the basis functions $\c ^{(n)}(p)$ and $\Y^{(n)}(p)$ recursively

\bea
\c ^{(n)}(p) &= &
\frac{1}{M_1}\, \Bigl ( \, \f^{(n)}_x(p)\, - \,\sum_{k=1}^{n-1} \c^{(k)}
(p)\, M_{n-k+1}\, \Bigr ) , \\
& &\nonumber \\
\Y^{(n)}(p) &= &\frac{1}{J_1}\,
\Bigl ( \, \f^{(n)}_y(p)\, - \,\sum_{k=1}^{n-1} \Y^{(k)}
(p)\, J_{n-k+1} \Bigr ) ,
\eea
where

\be
\f^{(n)}_x  (p) =  (p-x)^{-n-1/2}\, (p-y)^{-1/2}, \quad
\f^{(n)}_y  (p)  = (p-x)^{-1/2}\, (p-y)^{-n-1/2},
\ee{PhiXY}
following ref.\ \cite{Amb}. $\c ^{(n)}(p)$ and $\Y^{(n)}(p)$
are basis functions of
the linear operator $\widehat{V}^\prime$ appearing in the superloop equations
in the sense that

\bea
\Bigl ( \,
\widehat{V^\prime} - u_0(p) \, \Bigr ) \, \c ^{(n)}(p) &=& \frac{1}{(p-x)
^n}, \quad n\geq 1,  \label{VprimeChi} \\
& & \nonumber \\
\Bigl ( \,
\widehat{V^\prime} - u_0(p) \, \Bigr ) \, \Y^{(n)}(p) &=& \frac{1}{(p-y)
^n}, \quad n\geq 1. \label{VprimePsi}
\eea
Note that \mbox{$\f^{(0)}_x(p)=\f^{(0)}_y(p)\equiv \f^{(0)}(p)$} lies in the
kernel of $(\,\widehat{V^\prime} - u_0(p)\, )$ and constitutes the zero mode
of $v_g(p)$ mentioned above.

\subsection{The Planar Solution}

With these definitions at hand we may now state the planar ($g=0$) solution
of eqs.\ \refer{superloop1} and \refer{superloop2} obtained in refs.\
\cite{Alv2,Ple}:

\bea
W_0(\,\mid p) &=&\cint{\w}{C}\, \frac{V^\prime (\w )}{p-\w}\,
\biggl [ \, \frac{(p-x)(p-y)}{(\w -x)(\w -y)}\, \biggr ]^{1/2}\,
\phantom{+\,
\frac{1}{4}\, \frac{\X_2\, \XL}{M_1\, (x-y)}\, \f^{(1)}_x(p)} \label{W0p_0}\\
& &
+\,
\frac{1}{4}\, \frac{\X_2\, \XL}{M_1\, (x-y)}\, \f^{(1)}_x(p)
+\, \frac{1}{4}\, \frac{\L_2\, \XL}{J_1\, (x-y)}\, \f^{(1)}_y(p)\nonumber\\
&&\nonumber\\&&\nonumber\\
W_0(p\mid \, ) &=& \cint{\w}{C}\, \frac{\Y (\w )}{p-\w}\, \biggl [\,
\frac{(\w -x)(\w -y)}{(p-x)(p-y)}\, \biggr ] ^{1/2} \, +\, \frac{1}{2}\,
\frac{\X_1\, +\, \L_1}{[\, (p-x)(p-y)\, ]^{1/2}}.\nonumber \\ & &
\label{Wp0_0}
\eea

Here the endpoints  $x$ and $y$ of the cut on the real axis
are determined by the boundary conditions

\be
0=\cint{\w}{C}\, \frac {V^\prime(\w )}{\sqrt{(\w -x)(\w -y)}} , \quad \quad
1=\cint{\w}{C}\, \frac{\w\, V^\prime(\w )}{\sqrt{(\w -x)(\w -y)}},
\ee{determinexy}
deduced from our knowledge that $W(\, \mid p)= 1/p + {\cal O}(p^{-2})$.
This is also the reason why there are no zero mode contributions possible
for the \dol{W_g(\,\mid p)}.
We shall make use of the following rewriting of the purely bosonic part of
$W(\,\mid p)$ \cite{Amb3}

\be
u_0(p) = V^\prime (p) \, -\, \frac{1}{2}\, [(p-x)(p-y)]^{1/2}\,
\sum_{q=1}^\infty\,
\Bigl\{ (p-x)^{q-1}\, M_q\, +\, (p-y)^{q-1}\, J_q \Bigr\},
\ee{u0alternative}
derived by deforming the contour integral in eq.\ \refer{W0p_0} into one
surrounding the point $p$ and the other encircling infintity.
To take the residue at infinity one rewrites \dol{(p-\w)^{-1}} as

\be
\frac{1}{p-\w} = \frac{1}{2}  \frac{1}{(p-x)-(\w-x)}  + \frac{1}{2}
\frac{1}{(p-y)-(\w-y)},
\ee{trick}
and expands in \dol{\Bigl( \frac{p-x}{\w-x}\Bigr )}  and
\dol{\Bigl( \frac{p-y}{\w-y}\Bigr )} respectively. Doing this for
the fermionic \dol{W_0(p\mid \, )} yields

\be
v_0(p)=\Y (p) -
\, \frac{1}{2}\, [(p-x)(p-y)]^{-1/2}\, \sum_{q=2}^\infty\,
\Bigl\{ (p-x)^{q-1}\, \X_q\, +\, (p-y)^{q-1}\, \L_q \Bigr\}.
\ee{v0alternative}
It is important to realize that the bracketed terms in eq.\
\refer{u0alternative}
as well as in eq.\ \refer{v0alternative} are actually identical.
Here we see that the planar solution is special in the sense that it depends on
the full set of moments. Interestingly enough this is not the case for higher
genera.

\sect{The Double--Scaling Limit}

Similar to the situation in the hermitian matrix model the ``naive''
\mbox{$N\ra\infty$} continuum limit of the supereigenvalue model is
unsatisfactory as it leaves us only with the planar contributions, easily seen
from eqs.\
\refer{genusexpWnm} and \refer{genusexpF}. More interesting from the point
of view of continuum physics is the double--scaling limit. Crucial for this
scenario is the observation that there exists a subspace in the space of
couplings \mbox{$\{ g_k,\x_{k+1/2}\}$} where all higher genus contributions
to the free energy \mbox{$F_{g}$} diverge. This enables us to take the
double--scaling limit, where one simultaneously approaches the critical
subspace of couplings {\it and} takes \mbox{$N\ra \infty$}, giving
contributions
to the free energy from all genera. Let us see how this works in detail.

\subsection{Scaling of Moments and Basis Functions}

The analysis of the scaling behaviour for the bosonic quantities was carried
out
by Ambj\o rn {\it et al.\ }\cite{Amb,Amb3} in the framework
of the hermitian matrix model.
Consider the case of generic, i.e.\ non--symmetric, potentials $V(p)$ and $\Y
(p)$.
The $m$'th multicritical point is reached when the eigenvalue density
\mbox{$(u_0(p)-V^\prime (p))$} of eq.\ \refer{u0alternative}, which under
normal circumstances vanishes as a square root on both ends of its support,
aquires \dol{(m-1)} additional zeros at one end of the cut, say $x$. The
condition
for being at an $m$'th multicritical point then simply is

\be
M_1^c=M_2^c=\cdots M_{m-1}^c=0, \quad M_k^c\neq 0,\,\,\,\,  k\geq m,
\quad\quad J_l^c\neq 0, \,\,\,\, l\geq 1,
\ee{MJcritical}
defining a critical subspace in the space of bosonic couplings $g_k$.
Denote by \dol{g_k^c} a particular point in this subspace for which the
eigenvalue
distribution is confined to the interval \dol{[x_c,y_c]}. If we now move away
from this point the conditions of eq.\
\refer{MJcritical} will no longer be fulfilled and the cut will move to
the intervall \dol{[x,y]}. Assume we control this movement by the parameter $a$
and set \cite{Amb}

\bea
x &=& x_c - a\, \L^{1/m} \label{xscaling}\\
p &=& x_c + a\, \p \label{pscaling}
\eea
The scaling of $p$ and the introduction of its scaling variable $\p$ is
necessary in order to speak of the double--scaling limit of
the superloop correlators. $\L$ plays the role of the cosmological constant.
We will now deduce the scaling of $y$ by
further assuming that the critical subspace $\{g_k^c\}$ is reached as

\be
g_k = g\, \cdot\, g_k^c,
\ee{gscaling}
where \dol{g} is a function of $a$ to be determined. Imposing the boundary
conditions \refer{determinexy} yields

\be
y-y_c \sim a^m, \quad \mbox{and} \quad  g-1 \sim a^m.
\ee{ygscale}
Knowing this one easily computes the $m$'th multicritical scaling behaviour
of the bosonic moments

\be
M_k \sim a^{m-k}, \quad k\in [1,m-1],
\ee{Mscale}
while the higher $M$--moments and the $J$--moments do not scale.

Moreover the functions $\f^{(n)}_x(p)$  and $\f^{(n)}_y(p)$ are found to behave
like

\be
\f^{(n)}_x(p) \sim a^{-n-1/2}, \quad\quad \f^{(n)}_y(p) \sim a^{-1/2},
\ee{phiscaling}
from which one proves the scaling behaviour of the basis functions

\be
\c^{(n)}(p) \sim a^{-m-n+1/2},\quad\quad \Y^{(n)}(p) \sim a^{-1/2},
\ee{basisscaling}
following ref.\ \cite{Amb}.

Let us now turn to the scaling of the fermionic moments $\X_k$ and $\L_k$.
Similar to the bosonic case the function \dol{(v_0(p)-\Y (p))}  of eq.\
\refer{v0alternative} usually vanishes at the endpoints of the cut like a
square root.
We will fine tune the coupling constants \dol{\x_{k+1/2}} in such a manner
that \dol{(n-1)} extra zeros accumulate at $x$, i.e.\

\be
\X_2^c=\cdots \X_{n-1}^c=0,\quad \quad \X_k^c\neq 0,\,\,\,\,  k\geq n,
\quad \quad \L_l\neq 0, \,\,\,\, l\geq 2,
\ee{XLscale}
where \dol{\X_k^c \equiv \X_k [ x_c,y_c,\x^c_{k+1/2}]}. In addition the results
of
ref.\ \cite{Ple} tell us that the moments $\X_1$ and $\L_1$ will always
appear in the combination \dol{\XL}. This suggests to impose the
constraint \dol{\X_1^c-\L_1^c=0} on these moments. As there is no analogue to
the boundary conditions \refer{determinexy} for the fermionic quantities we
are free to choose the scaling of the coupling constants $\x_{k+1/2}$. We
set

\be
\x_{k+1/2} = a^{1/2}\, \, \x^c_{k+1/2},
\ee{xiscale}
and will comment on this choice later on.  From this one derives
\dol{(\X_1-\L_1) \sim a^{n-1/2}} and \dol{\X_k \sim a^{n-k+1/2}} for
\dol{k\in [2,n-1]}.
All other fermionic moments scale uniformly with $a^{1/2}$. So far the
fermionic
scaling is completely independent of the bosonic scaling, governed
by the integer $n$. We shall, however, introduce the requirement that the
scaling part of the bosonic one--superloop correlator of eq.\ \refer{W0p_0}
scales
uniformly, i.e.\ we require \dol{(u_0(p)-V^\prime(p))} and
\dol{\widehat{u}_0(p)}
to scale in the same way \cite{Alv2}. As \dol{(u_0(p)-V^\prime(p))\sim
a^{m-1/2}}
we arrive at the following condition on $n$:

\be
n = m+1.
\ee{nfix}
And therefore

\bea
\X_1-\L_1 \sim a^{m+1/2},&& \quad\quad \X_k \sim a^{m-k+3/2}, \quad
k\in [2,m]\nonumber\\ && \label{Xscale} \\
\X_k \sim a^{1/2}, \quad k>m &&\quad\quad \L_l\sim a^{1/2}, \quad l>
1.\nonumber
\eea
The double--scaling limit is now defined by letting \dol{N\ra\infty}
and \dol{a\ra 0} but keeping the string coupling constant
\dol{\a=a^{-2m-1}\, N^{-2}} fixed.
\footnote{Let us now comment on the choice of eq.\
\refer{xiscale}. At first sight one might have expected a scaling like
\dol{\x_{k+1/2}=[1+o(a)]\,\x_{k+1/2}^c}. This turns out to be inconsistent
because then the condition \refer{nfix} demands $n$ to be half--integer which
it
can not be.
If one takes the more general ansatz \dol{\x_{k+1/2}=
a^l\, \x_{k+1/2}^c} the condition of uniform scaling of \dol{W(\,\mid p)}
yields the allowed sequence
\dol{\{n,l\}=\{m+1,1/2\},\{m,3/2\},\{m-1,5/2\},\ldots}. The scaling of the
lowest moments in eq.\ \refer{Xscale} remains unchanged, the uniform scaling
however already starts with \dol{\X_{n}} scaling like $a^l$ and thus simply
reduces the number of double--scaling relevant terms.}

The above scaling behaviour may be shown to be equivalent to the one
obtained in ref.\ \cite{Alv2}. Based on these results Abdalla and Zadra
\cite{Zad} proved that the
central charge of the superconformal field theory
described in the continuum is given by

\be
\widehat{c}= 1\, -\, \frac{(2m-1)^2}{m} = 0, -\frac{7}{2}, \ldots \quad\quad
m\geq 1.
\ee{chat}
Similar to the situation in the bosonic model the \dol{m=1} fixed point may
not be reached by the described techniques and should be treated seperately.

\subsection{The Iteration for $u_g(p)$ and $v_g(p)$}

By making use of the above scaling properties of the moments and basis
functions
we may now develop the iterative procedure which allows us to calculate
directly the double--scaling relevant versions of $u_g(p)$ and
$v_g(p)$. The iterative scheme described in ref.\ \cite{Ple} is quite simple:
By eqs.\ \refer{VprimeChi} and \refer{VprimePsi} every $u_g$ and $v_g$ may be
written as a linear combination of basis functions $\c^{(n)}$ and $\Y^{(n)}$,
where the coefficients of this expansion are simply read off the poles at
$x$ and $y$ of the right hand sides of the superloop equations
\refer{order0genusg} and \refer{order1genusg}
after a partial fraction decomposition. To optimize the procedure to only
produce terms which are relevant in the double--scaling limit we have to
analyze the operators appearing on the right hand sides of the superloop
equations, i.e.\ the superloop insertion
operators $\d/\d V(p)$ and $\d/\d\Y(p)$ as well as \dol{(\widehat{\Y}-v_0(p))}.

{}From the point of view of the $a\ra 0$ limit the effect of a given operator
in
$\d/\d V(p)$ acting on an expression which scales with $a$ to some
power is to lower this power by a certain amount. Carefully examining each term
in $\d/\d V(p)$ shows that a is maximally lowered by a power of \dol{(m+3/2)}.
All operators which do not lower $a$ by this amount are subdominant
in the scaling limit and may be neglected. The outcome of this analysis for
$\d/\d V(p)$ is

\bea
\frac{\d}{\d V(p)_x} &=& \sum_{k=1}^\infty \frac{\d M_k}{\d V(p)_x}\,
\frac{\d}{\d M_k} \, + \, \frac{\d x}{\d V(p)}\, \frac{\d}{\d x} \nonumber\\
  & & \quad \, +\, \frac{\d\XL }{\d V(p)_x}\, \frac{\d}{\d \XL }
\, + \,  \sum_{k=2}^\infty \frac{\d\X_k}{\d V(p)_x}\,
\frac{\d}{\d\X_k} \label{dVx}
\eea
where \footnote{Here the subscript $x$ indicates that the critical
behaviour is associated with the endpoint $x$}

\be
\frac{\d M_k}{\d V(p)_x} \phantom{\X_1}
 = - (k+\frac{1}{2} )\, \f^{(k+1)}_x(p) \, +\, (k+\frac{1}{2})
\, \frac{M_{k+1}}{M_1}\,\f^{(1)}_x(p) \ee{dM}
\bea
\frac{\d x}{\d V(p)} \phantom{\X_1}
&=& \frac{1}{M_1}\,\f^{(1)}_x(p) \label{dx}\\ &&\nonumber\\
\frac{\d \XL }{\d V(p)_x} &=& \frac{1}{2}\, \frac{\X_2}{M_1}\, \f^{(1)}_x(p)
\label{dXL1} \\&&\nonumber \\
\frac{\d\X_k}{\d V(p)_x}\phantom{\X_1}
 &=& (k-\frac{1}{2})\, \frac{\X_{k+1}}{M_1}\, \f^{(1)}_x(p)
\label{dXk}
\eea
and

\be
\f^{(n)}_x(p) = (p-x)^{-n-1/2}\, {d_c}^{-1/2},
\ee{PhiXscale}
with \dol{d_c=x_c-y_c}. Repeating this analysis for the fermionic
superloop insertion operator $\d/\d\Y (p)$ shows that here $a$ is maximally
lowered by a power of \dol{(m+1)} and the relevant contributions are

\be
\frac{\d}{\d\Y (p)_x} = \frac{\d \XL }{\d\Y (p)_x}\, \frac{\d}{\d \XL } \, +\,
\sum_{k=2}^\infty \frac{\d\X_k}{\d\Y (p)_x}\, \frac{\d}{\d\X_k}
\ee{dYx}
with

\bea
\frac{\d \XL }{\d\Y (p)_x} &=& - d_c\, \f^{(0)}_x(p) \label{dYXL} \\
&&\nonumber\\
\frac{\d\X_k}{\d\Y (p)_x} \phantom{\,\X_1}
&=& - d_c \, \f^{(k-1)}_x(p). \label{dYXk}
\eea
Finally we state the double--scaling version of the operator
\dol{(\widehat{\Y}-v_0(p))} acting on the function $\f^{(n)}_x(p)$

\be
\Bigl( \widehat{\Y}-v_0(p)\Bigr ) _x\,
\f^{(n)}_x(p) = \sum_{k=1}^n \frac{\X_{n+2-k}}{d_c}\,
\frac{1}{(p-x)^k} \, + \, \frac{\XL }{2\, d_c}\, \frac{1}{(p-x)^{n+1}}.
\ee{YopPhi}
Here the operator \dol{(\widehat{\Y}-v_o(p))} is seen to increase the power of
$a$ of the expression it acts on by a factor of $m$.

We are now in a position to calculate the double--scaling limit of the right
hand
sides of the loop equations \refer{order0genusg} and \refer{order1genusg} for
$u_g(p)$ and $v_g(p)$ provided we know the double scaled versions of
\dol{u_1(p),\ldots, u_{g-1}(p)} and \dol{v_1(p),\ldots, v_{g-1}(p)}. As all
the $y$ dependence has disappeared we do not have to perform a decomposition
of the result. Moreover no $J_k$ and $\L_k$ dependent terms will contribute if
we
do not start out with any and we do not.

The starting point of the iterative procedure are of course the genus 0
correlators \cite{Ple}. Keeping only the doublescaling relevant
parts one has

\be
-\frac{d}{dq}\,\frac{\d v_0(q)}{\d\Y (p)_x}\, |_{p=q} =
\frac{\d u_0(p)}{\d V(p)_x}
= \frac{1}{8}\, \frac{1}{(p-x)^2}
\ee{u(p,p|)}
and

\be
\frac{\d v_0(p)}{\d V(p)_x} = \Bigl [ - \frac{\XL}{4 \, d_c\, M_1}\Bigr ]\,
\frac{1}
{(p-x)^3}\, + \, \Bigl [ \frac{\X_2}{4\, d_c\, M_1}\Bigr ]\,\frac{1}{(p-x)^2}.
\ee{W_0(p|p)}
The higher genus correlators $u_g(p)$ and $v_g(p)$ are expressed as linear
combinations of the basis function $\c^{(n)}(p)$ and take the general form
\cite{Ple}

\be
u_g(p)= \sum_{k=1}^{3g-1}\, A^{(k)}_g \, \c^{(k)}(p) \quad \mbox{and}\quad
v_g(p)= \sum_{k=1}^{3g}\, B^{(k)}_g \, \c^{(k)}(p)\, + \, \k_g\, \f^{(0)}(p),
\ee{uvgeneral}
where $\k_g$ is the zero mode coefficient not determined by the
first two superloop
equations \refer{order0genusg} and \refer{order1genusg}. Note
that the \dol{A^{(k)}_g}
coefficients should up to a factor of two be identical to those of the
hermitian
matrix model obtained in ref.\ \cite{Amb}.
We have calculated the $A^{(k)}_g$ and $B^{(k)}_g$
coefficients in the double--scaling limit for \dol{g=1,2} and $3$ with the aid
of {\it Maple}. The results of Ambj\o rn {\it et al.\ }\cite{Amb}
for the $A^{(k)}_g$ coefficients could be reproduced.

Before we state our explicit results let us turn to the general
scaling behaviour of the one--superloop correlators

\be
W_g(\,\mid p) \sim a^{(1-2g)\, (m+1/2)\, -\, 1}, \quad
\mbox{and}\quad
W_g(p\mid \, ) \sim a^{(1-2g)\, (m+1/2)\, -\, 1/2},
\ee{Wscaling}
which one proves by induction. The structure of the
coefficients $B_g^{(k)}$ is seen to be

\be
B_g^{(k)}= \sum_{\a_j, \b}\, \langle \a_1,\ldots , \a_s, \b \mid \a
\rangle_{g,k}
\,\frac{M_{\a_1} \ldots M_{\a_s}\, \X_\b}{{M_1}^\a\, {d_c}^g},
\ee{Bgkgeneral}
where the brackets denote rational numbers and where we write $\X_1$ for
\dol{\XL}. One shows that the
$\a$, $\a_j$, $\b$ and $s$ obey the conditions

\be
\a=2g+s-1, \quad\mbox{and} \quad \sum_{j=1}^s (\a_j -1) = 3g+1-\b-k
\ee{Bgkcond}
with \dol{\a_j\in [2,3g]} and \dol{\b\in [1,3g]}.
For the zero mode coefficient $\k_g$ the general structure is given by the same
expansion as eq.\ \refer{Bgkgeneral} with \dol{k=0}.
The conditions on $\a$, $\a_j$, $\b$ and $s$ then read

\be
\a=2g+s, \quad\mbox{and} \quad \sum_{j=1}^s (\a_j -1) = 3g+1-\b
\ee{kappacond}
where \dol{\a_j\in [2,3g]} and \dol{\b\in [1,3g+1]}. Similar relations hold
for the \dol{A_g^{(k)}} \cite{Amb}.

The explicit results for the $B^{(k)}_g$
coefficients for \dol{g=1} and \dol{g=2} now read

\bea
B_1^{(1)} &=&-{\frac {\X_{{3}}}{8\,M_{{1}}\,d_c}}+{\frac {M_{{2}}\X_{{2
}}}{8\,{M_{{1}}}^{2}\, d_c}},
\nonumber \\
B_1^{(2)}&=&{\frac {M_{{2}}\,\XL}{16\,{
M_{{1}}}^{2}d_c}}+{\frac {\X_{{2}}}{8\,M_{{1}}\, d_c}},
\quad\quad
B_1^{(3)}=-{\frac {5\,{\XL}}{16\,M_{{1}}\, d_c}}.
\nonumber \\&&
\nonumber \\
B_2^{(1)}&=&{\frac {203\,M_{{2}}\X_{{5}}}{128\,{d_c}^{2}{M_{{1}}}^{
4}}}-{\frac {145\,{M_{{3}}}^{2}\X_{{2}}}{128\,{M_{{1}}}^{5}{d_c}^{2}}}-{
\frac {105\,\X_{{6}}}{128\,{d_c}^{2}{M_{{1}}}^{3}}}+{\frac {63\,{M_{{2}}}
^{3}\X_{{3}}}{32\,{M_{{1}}}^{6}{d_c}^{2}}} \nonumber\\
 &&+{\frac {105\,M_{{4}}\X_{{3}}}{
128\,{d_c}^{2}{M_{{1}}}^{4}}}+{\frac {145\,M_{{3}}\X_{{4}}}{128\,{d_c}^{2}{
M_{{1}}}^{4}}}+{\frac {105\,M_{{5}}\X_{{2}}}{128\,{d_c}^{2}{M_{{1}}}^{4}}
}-{\frac {77\,M_{{4}}M_{{2}}\X_{{2}}}{32\,{M_{{1}}}^{5}{d_c}^{2}}}\nonumber\\
&&
-{\frac {87\,M_{{3}}M_{{2}}\X_{{3}}}{32\,{M_{{1}}}^{5}{d_c}^{2}}}
+{\frac {
75\,M_{{3}}{M_{{2}}}^{2}\X_{{2}}}{16\,{M_{{1}}}^{6}{d_c}^{2}}}-{\frac {63
\,{M_{{2}}}^{4}\X_{{2}}}{32\,{M_{{1}}}^{7}{d_c}^{2}}}-{\frac {63\,{M_{{2}
}}^{2}\X_{{4}}}{32\,{M_{{1}}}^{5}{d_c}^{2}}},\nonumber \\
B_2^{(2)}&=&-{\frac {21\,{M_{{
2}}}^{3}\X_{{2}}}{64\,{M_{{1}}}^{6}{d_c}^{2}}}-{\frac {21\,M_{{2}}\X_{{4}}
}{64\,{d_c}^{2}{M_{{1}}}^{4}}}+{\frac {77\,M_{{3}}M_{{2}}\X_{{2}}}{128\,{
M_{{1}}}^{5}{d_c}^{2}}}-{\frac {35\,M_{{4}}\X_{{2}}}{128\,{d_c}^{2}{M_{{1}}
}^{4}}} \nonumber \\
&&+{\frac {75\,M_{{3}}{M_{{2}}}^{2}{\XL}}{32\,{M_{{1}}}^
{6}{d_c}^{2}}}+{\frac {105\,M_{{5}}{\XL}}{256\,{d_c}^{2}{M_{{1}}}
^{4}}}-{\frac {63\,{M_{{2}}}^{4}{\XL}}{64\,{M_{{1}}}^{7}{d_c}^{
2}}} \nonumber \\
&&+{\frac {35\,\X_{{5}}}{128\,{M_{{1}}}^{3}{d_c}^{2}}}-{\frac {145\,{M_{
{3}}}^{2}{\XL}}{256\,{M_{{1}}}^{5}{d_c}^{2}}}+{\frac {21\,{M_{{
2}}}^{2}\X_{{3}}}{64\,{M_{{1}}}^{5}{d_c}^{2}}}-{\frac {35\,M_{{3}}\X_{{3}}
}{128\,{d_c}^{2}{M_{{1}}}^{4}}}, \nonumber \\
&& -{\frac {77\,M_{{4}}M_{{2}}{\XL}}{64\,{M_{{1}}}^{5}{d_c}^{2
}}} \nonumber\\
B_2^{(3)}&=&-{\frac {599\,M_{{3}}M_{{2}}
{\XL}}{128\,{M_{{1}}}^{5}{d_c}^{2}}}+{\frac {105\,{M_{{2}}}^{3}
{\XL}}{32\,{M_{{1}}}^{6}{d_c}^{2}}}-{\frac {5\,M_{{3}}\X_{{2}}}{
16\,{d_c}^{2}{M_{{1}}}^{4}}} \nonumber \\
&& +{\frac {385\,M_{{4}}{\XL}}{256\,{d_c
}^{2}{M_{{1}}}^{4}}} +{\frac {21\,{M_{{2}}}^{2}\X_{{2}}}{64\,{M_{{1}}}^{
5}{d_c}^{2}}}+{\frac {7\,M_{{2}}\X_{{3}}}{128\,{d_c}^{2}{M_{{1}}}^{4}}},
\nonumber \\
B_2^{(4)}&=&-{\frac {357\,{M_{{2}}}^{2}{\XL}}{64\,{M_{{1}}}^{5}{
d_c}^{2}}}+{\frac {875\,M_{{3}}{\XL}}{256\,{d_c}^{2}{M_{{1}}}^{4}
}}-{\frac {35\,\X_{{3}}}{128\,{M_{{1}}}^{3}{d_c}^{2}}}-{\frac {63\,M_{{2}
}\X_{{2}}}{128\,{d_c}^{2}{M_{{1}}}^{4}}},\nonumber \\
B_2^{(5)}&=&{\frac {105\,\X_{{2}}}{
128\,{M_{{1}}}^{3}{d_c}^{2}}}+{\frac {1617\,M_{{2}}{\XL}}{256\,
{d_c}^{2}{M_{{1}}}^{4}}},
\quad\quad
B_2^{(6)}=-{\frac {1155\,{\XL}}{256\,{
M_{{1}}}^{3}{d_c}^{2}}}. \label{Bees}
\eea
Note that the terms listed above are only potentially relevant, depending on
which multi--critical model one wishes to consider. For an $m$'th
multi--critical
model all terms containing $M_k$, \dol{k>m}, or $\X_l$, \dol{l>m+1}, vanish in
the double--scaling limit. We remind the reader that we assumed to have a
non--symmetric potential and that the critical behaviour was associated with
the endpoint $x$. In the case where the critical behaviour is associated with
the
endpoint $y$ all formulas in this section still hold provided $d_c$ is replaced
by $-d_c$, $M_k$ by $J_k$, $\X_k$ by $\L_k$ and $x$ by $y$.

\subsection{The Iteration for $F_g$, $\k_g$ and $\widehat{u}_g(p)$}

Having computed $u_g(p)$ and $v_g(p)$ (up to the zero mode) we may now
proceed to calculate the free energy $F_g$ and the zero mode coefficient
$\k_g$.
This is done by rewriting $u_g$ and $v_g$ as total derivatives in the superloop
insertion operators \dol{\d/\d V(p)} and \dol{\d/\d\Y (p)} respectively,
yielding
the bosonic and doubly fermionic parts of $F_g$ as well as $\k_g$. The
procedure
to compute the bosonic part of the free energy \dol{\Fbg}
 works just as in the hermitian
matrix model described in ref.\ \cite{Amb}. We have \dol{\Fbg=2\,
F^{\mbox{\scriptsize
herm}}_g.}

{}From eq.\ \refer{Wscaling} we see that \dol{F_g} scales as

\be
F_g= \Fbg \, +\, \Ffg\, \sim a^{(2-2g)(m+1/2)},
\ee{Fscaling}
just as the hermitian matrix model at its $m$'th multicritical point.

To obtain the doubly femionic part \dol{\Ffg} simply
rewrite the basis functions $\c^{(n)}(p)$ of $v_g(p)$ appearing in
\refer{uvgeneral}
in terms of the functions \dol{\f^{(n)}_x(p)} which by eqs.\ \refer{dYXL} and
\refer{dYXk} are nothing but total derivatives in \dol{\d/\d\Y (p)}. Doing this
for $v_g(p)$ allows one to directly deduce the form of \dol{\Ffg} and $\k_g$.
\footnote{Actually the presentation here is somewhat misleading. In order to
compute the $u_g(p)$ and $v_g(p)$ by iteration in the way outlined in the
previous subsection one needs to know the {\it full} \dol{v_1(p), \ldots,
v_{g-1}(p)}
including the zero mode coefficients \dol{\k_1, \ldots, \k_{g-1}}. In
practice one hence computes all quantities at genus $g$, i.e.\ $u_g$, $v_g$,
$\k_g$
and $F_g$, before proceeding to genus \dol{(g+1)}.}

The explicit results for genus one are

\be
F^{\mbox{\scriptsize ferm}}_1=
\XL\Bigl (-{\frac {
5\,{\X}_{{4}}}{16\,{d_c}^{2}{M_{{1}}}^{2}}}+{\frac {3\,M_{{2}}\, {\X}_{{3}}}
{8\,{d_c}^{2}{M_{{1}}}^{3}}}+{\frac {5\,M_{{3}}\, {\X}_{{2}}}
{16\,{d_c}^{2}{M_{{1}}}^{3}}}-{\frac {3\,{M_{{2}}}^{2}\, {\X}_{{2}}}{8
\,{d_c}^{2}{M_{{1}}}^{4}}}\Bigr )
+{\frac {\X_{{2}}\, {\X}_{{3}}}{8\,{d_c}^{2}{M_{{1}}}^{2}}}
\ee{F1}
and

\be
\k_1={\frac {5\,\X_{{4}}}{16\,{M_{{1}}}^{2}
d_c}}-{\frac {3\,M_{{2}}\,\X_{{3}}}{8\,{M_{{1}}}^{3}d_c}}-{\frac
{5\,M_{{3}}\,\X
_{{2}}}{16\,{M_{{1}}}^{3}d_c}}+{\frac {3\,{M_{{2}}}^{2}\,\X_{{2}}}{8\,{M_{{
1}}}^{4}d_c}}.
\ee{}

This expression may of course be alternatively obtained by taking the double
scaling limit of the genus one result of ref.\ \cite{Ple} computed
away from the scaling limit.

Before presenting explicit results for genus two, let us describe the
general structure of \dol{\Ffg} for \dol{g\geq 1}

\be
\Ffg= \sum _{\a_j, \b_i}\, \langle \a_1,\ldots , \a_s, \b_1,\b_2\mid \a\rangle
_g
\, \frac{\X_{\b_1}\, \X_{\b_2}\, M_{\a_1} \ldots M_{\a_s}}{{M_1}^\a\,
{d_c}^{g+1}},
\ee{Ffggeneral}
where we write \dol{\X_1} for \dol{\XL} and where the brackets denote
rational numbers as before. The \dol{\a_j}, $\b_i$, $\a$ and $s$ are
subject to the constraints

\be
\a=2g+s, \quad \mbox{and}\quad \sum_{j=1}^s (\a_j -1)= 3g+2-\b_1-\b_2,
\ee{alphaconstraints}
where \dol{\a_i\in [2,3g]} and \dol{k_i\in [1,3g+1]}.
Similar relations hold for $\Fbg$ \cite{Amb}.

The result for genus two reads

\bea
F_2^{\mbox{\scriptsize ferm}}&=&
\XL \, \biggl[ \,
 {\frac {1015\,\X_{{5}}\, M_{{3}}}{128\,{M_{{1}}}^{5}{d_c}^{3}}}
-{\frac {375\,{\X}_{{4}}\, M_{{2}}\, M_{{3}}}
   {16\,{M_{{1}}}^{6}{d_c}^{3}}}                                \nonumber\\&&
+{\frac {315\,\X_{{4}}\, {M_{{2}}}^{3}}{16\,{M_{{1}}}^{7}
   {d_c}^{3}}}
+{\frac {385\,\X_{{4}}\, M_{{4}}}{64\,{M_{{1}}}^{5}{d_c}^{3}}}
+{\frac {1323\,\X_{{2}}\, {M_{{2}}}^{5}}
   {64\,{M_{{1}}}^{9}{d_c}^{3}}}
+{\frac {693\, \X_{{6}}\, M_{{2}}}{64\,{M_{{1}}}^{5}{d_c}^{3}}} \nonumber\\&&
-{\frac {1155\,\X_{{7}}}{256\,{M_{{1}}}^{4}{d_c}^{3}}}
-{\frac {63\,\X_{{2}}\, M_{{2}}\, M_{{5}}}
   {4\,{M_{{1}}}^{6}{d_c}^{3}}}
-{\frac {525\,\X_{{5}}\, {M_{{2}}}^{2}}
   {32\,{M_{{1}}}^{6}{d_c}^{3}}}
-{\frac {1155\,\X_{{3}}\, M_{{2}}\, M_{{4}}}
  {64\,{M_{{1}}}^{6}{d_c}^{3}}}                \nonumber\\&&
+{\frac {315\,\X_{{3}}\, M_{{5}}}{64\,{M_{{1}}}^{5}{d_c}^{3}}}
-{\frac {2175\,\X_{{3}}\,
  {M_{{3}}}^{2}}{256\,{M_{{1}}}^{6}{d_c}^{3}}}
+{\frac {675\,\X_{{3}}\, {M_{{2}}}^{2}\, M_{{3}}}
  {16\,{M_{{1}}}^{7}{d_c}^{3}}}
-{\frac {1785\,\X_{{2}}\, M_{{3}}\, M_{{4}}}
  {128\,{M_{{1}}}^{6}{d_c}^{3}}}                         \nonumber\\&&
-{\frac {495\,\X_{{2}}\, {M_{{2}}}^{3}\, M_{{3}}}
  {8\,{M_{{1}}}^{8}{d_c}^{3}}}
+{\frac {2205\,\X_{{2}}\, {M_{{2}}}^{2}\, M_{{4}}}
  {64\,{M_{{1}}}^{7}{d_c}^{3}}}
-{\frac {1323\,\X_{{3}}\, {M_{{2}}}^{4}}
  {64\,{M_{{1}}}^{8}{d_c}^{3}}}
+{\frac {8175\,\X_{{2}}\, {M_{{3}}}^{2}M_{{2}}}
  {256\,{M_{{1}}}^{7}{d_c}^{3}}}     \nonumber\\ & &
+{\frac {1155\,\X_{{2}}\, M_{{6}}}
  {256\,{M_{{1}}}^{5}{d_c}^{3}}}
\, \biggr]
-{\frac {21\,\X_{{2}}\, {\X}_{{5}}\, M_{{2}}}{16\,{M_{{1}}}^{5}{d_c}^{3}}}
+{\frac {105\,\X_{{2}}\, {\X}_{{6}}}{128\, M_{{1}}^{4}{d_c}^{3}}}
-{\frac {145\,\X_{{2}}\, {\X}_{{4}}\, M_{{3}}}{128\,{M_{{1}}}^{5}{d_c}^{3}}}
\nonumber \\ &&
+{\frac {105\,\X_{{2}}\, {\X}_{{4}}{M_{{2}}}^{2}}{64\,{M_{{1}}}^{6}{d_c}^{3}}}
+{\frac {21\,{\X}_{{3}}\, \X_{{4}}\, M_{{2}}}{64\,{M_{{1}}}^{5}{d_c}^{3}}}
-{\frac {35\,\X_{{3}}\, {\X}_{{5}}}{128\,{M_{{1}}}^{4}{d_c}^{3}}}
-{\frac {63\,\X_{{2}}\, {\X}_{{3}}\,
{M_{{2}}}^{3}}{32\,{M_{{1}}}^{7}{d_c}^{3}}}
\nonumber\\ &&
-{\frac {35\,\X_{{2}}\, {\X}_{{3}}\, M_{{4}}}{32\,{M_{{1}}}^{5}{d_c}^{3}}}
+{\frac {195\,\X_{{2}}\, {\X}_{{3}}\, M_{{2}}\, M_{{3}}}
  {64\,{M_{{1}}}^{6}{d_c}^{3}}}.  \label{Ff2}
\eea
Of course so far we have determined only the coefficients $F_g$ of the genus
expansion of the free energy (cf.\ eq.\ \refer{genusexpF}).
 For an $m$'th multicritical model the relevant expansion parameter in the
 double--scaling limit is the string coupling constant \dol{\a=a^{-2m-1}\,
N^{-2}}.
 If we introduce the bosonic and fermionic scaling moments $\m_k$ and $\t_k$
 by (cf.\ eqs.\ \refer{Mscale} and \refer{Xscale})

 \be
 M_k=a^{m-k}\, \m_k, \quad k\in [1,m],
 \ee{scalingmoments}
 $$
 \XL= a^{m+1/2}\, \t_1, \quad\quad \X_l=a^{m-l+3/2}\,\t_l,\quad
 l\in [2,m+1],
 $$
 we get by replacing $M_k$ by $\m_k$ and setting $M_k$ equal to zero for
\dol{k>m}
 as well as replacing $\X_l$ by $\t_l$ and setting $\X_l$ to zero for
\dol{l>m+1}
 in the formulas above exactly the coefficients of the expansion in the string
 coupling constant.
Needless to say that these results apply for non--symmetric potentials
where the critical behaviour is associated with the endpoint $y$ as well by
performing the usual replacements.

For the zero mode coefficient at genus two we find

\bea
{\k_2}&=&{\frac {495\,\X_{{2}}M_{{3}}{M_{{2}}}
^{3}}{8\,{d_c}^{2}{M_{{1}}}^{8}}}+{\frac {63\,\X_{{2}}M_{{5}}M_{{2}}}{4\,
{d_c}^{2}{M_{{1}}}^{6}}}-{\frac {315\,\X_{{3}}M_{{5}}}{64\,{d_c}^{2}{M_{{1}
}}^{5}}}+{\frac {1323\,\X_{{3}}{M_{{2}}}^{4}}{64\,{d_c}^{2}{M_{{1}}}^{8}}
}
\nonumber\\&&
-{\frac {675\,\X_{{3}}M_{{3}}{M_{{2}}}^{2}}{16\,{d_c}^{2}{M_{{1}}}^{7}}}
+{\frac {2175\,\X_{{3}}{M_{{3}}}^{2}}{256\,{d_c}^{2}{M_{{1}}}^{6}}}-{
\frac {8175\,\X_{{2}}{M_{{3}}}^{2}M_{{2}}}{256\,{d_c}^{2}{M_{{1}}}^{7}}}-
{\frac {385\,\X_{{4}}M_{{4}}}{64\,{d_c}^{2}{M_{{1}}}^{5}}}
\nonumber\\&&
-{\frac {2205\,
\X_{{2}}M_{{4}}{M_{{2}}}^{2}}{64\,{d_c}^{2}{M_{{1}}}^{7}}}+{\frac {375\,\X
_{{4}}M_{{3}}M_{{2}}}{16\,{d_c}^{2}{M_{{1}}}^{6}}}-{\frac {1155\,\X_{{2}}
M_{{6}}}{256\,{d_c}^{2}{M_{{1}}}^{5}}}+{\frac {1785\,\X_{{2}}M_{{4}}M_{{3
}}}{128\,{d_c}^{2}{M_{{1}}}^{6}}}
\nonumber\\&&
-{\frac {693\,\X_{{6}}\, M_{{2}}}{64\,{d_c}^{
2}{M_{{1}}}^{5}}}-{\frac {315\,\X_{{4}}{M_{{2}}}^{3}}{16\,{d_c}^{2}{M_{{1
}}}^{7}}}+{\frac {525\,\X_{{5}}{M_{{2}}}^{2}}{32\,{d_c}^{2}{M_{{1}}}^{6}}
}-{\frac {1015\,\X_{{5}}M_{{3}}}{128\,{d_c}^{2}{M_{{1}}}^{5}}}
\nonumber\\&&
-{\frac {
1323\,\X_{{2}}{M_{{2}}}^{5}}{64\,{d_c}^{2}{M_{{1}}}^{9}}}+{\frac {1155\,\X
_{{3}}M_{{4}}M_{{2}}}{64\,{d_c}^{2}{M_{{1}}}^{6}}}+{\frac {1155\,\X_{{7}}
}{256\,{d_c}^{2}{M_{{1}}}^{4}}}. \label{kappa2}
\eea

We obtained these results with the aid of a {\it Maple} program which performs
the iteration up to arbitrary genus.
\footnote{The reader interested in higher--genus results should feel free
to contact the author.}
In practice the expressions become quite
lengthy, e.g.\ \dol{F^{\mbox{\scriptsize ferm}}_3} consists of 114 terms.

For the knowledge of the full \dol{W_g(\,\mid p)} it remains to compute
$\widehat{u}_g(p)$. This is of course done by applying \dol{\d/\d V(p)} to
\dol{\Ffg}. The general structure of $\widehat{u}_g(p)$ turns out to be
\cite{Ple}

\be
\widehat{u}_g(p)= \sum_{k=1}^{3g+1} \widehat{A}_g^{(k)}\, \c^{(k)}(p),
\ee{uhatgeneral}
where

\be
\widehat{A}_g^{(k)}= \sum_{\a_j, \b_1,\b_2}\,
\langle \a_1,\ldots ,\a_s,\b_1,\b_2\mid \a\rangle_{g,k}\,
\frac{M_{\a_1}\ldots M_{\a_s}\, \X_{\b_1}\,\X_{\b_2}}{{M_1}^\a\, {d_c}^{g+1}}
\ee{Ahatgkgeneral}
underlying the conditions

\be
\a= s+3g, \quad\mbox{and}\quad \sum_{j=1}^s (\a_j-1)= 4+3g-k-\b_1-\b_2,
\ee{Ahatconds}
with \dol{\a_j\in [2,3g+1]} and \dol{\b_i\in [1,3g+2]}. Due to space let us
only
state the genus one results

\bea
\widehat{A}_1^{(1)}&=&{\frac {5\,\X_{{2}}\,\X_{{4}}}{32\,{d_c}^{2}{M_{{1}}}^{2}
}}+{\frac {3\,M_{{2}}\X_{{2}}{\X}_{{3}}}{16\,{d_c}^{2}{M_{{1}}}^{3}}}
-{\frac {35\,{\XL}{\X}_{{5}}}{32\,{d_c}^{2}{M_{{1}}}^{2}}
} \nonumber \\&&
+{\frac {15\, M_{{3}}{\XL}\,{\X}_{{3}}}{32\,{d_c}^{2}{M_{{1}}
}^{3}}}
+{\frac {15\, M_{{2}}{\XL}\,{\X}_{{4}}}{16\,{d_c}^{2}{M
_{{1}}}^{3}}}-{\frac {9\,{M_{{2}}}^{2}{\XL}\,{\X}_{{3}}}{16
\,{d_c}^{2}{M_{{1}}}^{4}}},
\nonumber \\
\widehat{A}_1^{(2)}&=&-{\frac {15\,{\XL}\,{\X}_{{4}}}
{16\,{d_c}^{2}{M_{{1}}}^{2}}}+{\frac {3\,\X_{{2}}\, {\X}_{{3}}
}{8\,{d_c}^{2}{M_{{1}}}^{2}}}+{\frac {3\,M_{{2}}{\XL}\, {\X}_{
{3}}}{4\,{d_c}^{2}{M_{{1}}}^{3}}}\nonumber \\&&
+{\frac {5\,M_{{3}}{\XL}{\X}_{{2}}}
{16\,{d_c}^{2}{M_{{1}}}^{3}}}-{\frac {3\,{M_{{2}
}}^{2}{\XL}\, {\X}_{{2}}}{8\,{d_c}^{2}{M_{{1}}}^{4}}}, \label{uhat1}\\
\widehat{A}_1^{(3)}&=&-{\frac {
15\,{\XL}{\X}_{{3}}}{16\,{d_c}^{2}{M_{{1}}}^{2}}}+{\frac {
25\, M_{{2}}\, {\XL}\, {\X}_{{2}}}{32\,{d_c}^{2}{M_{{1}}}^{3}}},
\quad\quad
\widehat{A}_1^{(4)}=-{\frac {35\,{\XL}\, {\X}_{{2}}}{32\,{d_c}^{2}{M_{
{1}}}^{2}}}. \nonumber
\eea
This concludes our analysis of the double--scaling limit for generic
potentials.

The case of symmetric bosonic and generic fermionic potentials was considered
in ref.\ \cite{Alv} where a doubling of degrees of freedom was observed for
genus
zero. With the methods presented in this paper one can see that this holds for
higher genera as well, i.e. the free energy here takes the form

\be
F_g^{\mbox{\scriptsize symm}}= 2\, \Fbg \, + \, F_g^{\mbox{\scriptsize ferm,x}}
\, +\, F_g^{\mbox{\scriptsize ferm,y}},
\ee{Fgsymm}
where $\Fbg$ denotes the bosonic part of the free energy for generic
potentials,
\dol{F_g^{\mbox{\scriptsize ferm,x}}} and
\dol{F_g^{\mbox{\scriptsize ferm,y}}} denote the doubly fermionic parts
in the generic case
where the critical behaviour is associated with the endpoints $x$ and
 \dol{y=-x}
respectively (cf.\ eq.\ \refer{Ffggeneral}). If one chooses to take
the fermionic potential
to be symmetric as well the doubly fermionic part of the free energy will
vanish, which may  be directly deduced from the results of ref.\ \cite{Bec}.

\sect{Conclusions}

We have studied the double scaled supereigenvalue model in the moment
description. The $m$'th multi--critical point was identified and the scaling
properties of the moments and basis functions derived. The iterative scheme
for the calculation of higher--genus contributions could
be optimized to produce
only terms relevant in the double--scaling limit. The general form of the
free energy and the one--superloop correlators at genus $g$ were found.
We presented explicit results up to genus two.

We believe that this paper shows once more the effectiveness of the iterative
scheme and the moment description of Ambj\o rn {\it et al.\ } The analogy of
structures in the hermitian one matrix model and in the supereigenvalue
model continues to hold. It is interesting to note that the analogous
 computations
for the hermitian matrix model \cite{Amb} were the basis of a transition
from the double scaled one--matrix to the Kontsevich model
\cite{Amb3}. One might speculate that our results are connected to properties
of the moduli space of super--Riemann surfaces.

There are more interesting
unanswered questions to be adressed,
such as the supersymmetric generalization of two
and multi--matrix models, where one would hope to reach new minimal
superconformal models, or the generalization of matrix models in external
fields,
perhaps a step towards a matrix based formulation of the supereigenvalue model.

\bigskip

\underline{Acknowledgements:} I am particularly grateful to E.\ Abdalla
and A.\ Zadra for pointing out the situation at the \dol{m=1} fixed--point
to me.
Moreover I wish to thank P.~Adamietz, G.~Akemann, J.~Bischoff
and O.~Lechtenfeld for interesting and valuable discussions.

\pagebreak

\end{document}